\documentclass[prc,twocolumn,showpacs,floatfix,superscriptaddress,nofootinbib]{revtex4}
\usepackage{graphicx}
\usepackage{amssymb}
\usepackage{dcolumn}
\usepackage{amsmath}
\usepackage{bm}
\usepackage{textcomp}
\usepackage{epstopdf}
\usepackage{color}
\usepackage{ulem}
\usepackage[toc,page,title,titletoc,header]{appendix}
\usepackage[colorlinks,
            linkcolor=blue,
            anchorcolor=blue,
            citecolor=blue
            ]{hyperref}
\usepackage{txfonts}
\usepackage{caption}

\begin{document}

\title{Dynamical critical fluctuations near the QCD critical point with hydrodynamic cooling rate}

\begin{abstract}

Within the model A in the Hohenberg's dynamical universality classification, we investigate the critical slowing down effects on the critical fluctuations driven by the expanding quark-gluon plasma, using a trajectory and cooling rate obtained from hydrodynamics. We numerically solved the Langevin dynamics of the non-conserved order parameter field and find that, compared with commonly used Hubble-like expansion, the cooling rate of a realistic hydrodynamic system is pretty large and the associated critical slowing down effects strongly suppress the higher-order cumulants of the order parameter field ({\it e.g.,} $C_4$). Furthermore, for an evolving system that approaches the critical point, such critical slowing down suppression overcomes the enhancement of the critical fluctuations, which indicates that the largest fluctuations of the order parameter field ({\it i.e.,} $C_2$)  do not necessarily associate with the evolving trajectory closest to the critical point.
\end{abstract}
\author{Shian Tang}
\email{tangshian@pku.edu.cn}
\affiliation{School of Physics, Peking University, Beijing 100871, China}

\author{Shanjin Wu}
\email{shanjinwu2014@pku.edu.cn}
\affiliation{School of Nuclear Science and Technology, Lanzhou University, Lanzhou 730000, China}
\affiliation{Center for High Energy Physics, Peking University, Beijing 100871, China}
\affiliation{School of Physics, Peking University, Beijing 100871, China}
\affiliation{Collaborative Innovation Center of Quantum Matter, Beijing 100871, China}

\author{Huichao Song}
\email{huichaosong@pku.edu.cn}
\affiliation{School of Physics, Peking University, Beijing 100871, China}
\affiliation{Center for High Energy Physics, Peking University, Beijing 100871, China}
\affiliation{Collaborative Innovation Center of Quantum Matter, Beijing 100871, China}

\maketitle

\section{Introduction}\label{Introduction}
The quantum chromodynamics(QCD) phase diagram is one of the most important topics in high-energy nuclear physics. Lattice QCD calculations show that the phase transition between hadron gas and quark-gluon plasma is a crossover at the vanishing baryon chemical potential~\cite{Aoki:2006we,Ding:2015ona,Bazavov:2019lgz,Ratti:2018ksb} and the QCD-based effective models predict a first-order phase transition in the finite baryon chemical potential region~\cite{Fukushima:2010bq,Fukushima:2013rx,Fischer:2018sdj,Fu:2022gou}, which suggests a critical end-point on the phase diagram~\cite{Stephanov:1998dy,Stephanov:2004wx,Stephanov:2006zvm,Asakawa:2015ybt}. However, the locations of the critical point predicted by the effective models are parameter-dependent and the lattice QCD simulations suffer from a sign problem in the region of finite baryon chemical potential~\cite{Hands:2007by}. The Beam Energy Scan (BES) program at the Relativistic Heavy Ion Collider(RHIC) aims to search for the critical point by scanning the QCD phase diagram~\cite{STAR:2010vob,Luo:2017faz,Bzdak:2019pkr}. Its first phase BES-I has measured the cumulants of the net-proton, net-charge, and net-kaon multiplicity distributions in Au+Au collisions with the collision energies ranging from 7 to 200 GeV~\cite{STAR:2010mib,STAR:2013gus,Luo:2015ewa,Luo:2015doi,STAR:2014egu,Thader:2016gpa,STAR:2020tga,STAR:2021iop}, and the second phase BES-II in progress will provide measurements with even higher statistics.

Theoretically, one of the most distinctive features of a system at the critical point is the divergence of the correlation length, which leads to several striking properties, such as large fluctuations, singularity, universality, and the critical slowing down effect. In relativistic heavy-ion collisions, several thermodynamic quantities are expected to fluctuate strongly near the critical point~\cite{Stephanov:1998dy,Stephanov:1999zu}, which could be imprinted in the associated experimental measurements~\cite{Hatta:2003wn,Kitazawa:2012at,Kitazawa:2011wh,Sun:2018jhg,Sun:2017xrx,
Shuryak:2018lgd,Wu:2022cbh,Wu:2022fuz}. For the net-proton multiplicity fluctuations, a non-monotonic behavior of the kurtosis as a function of  the collision energy was predicted~\cite{Kitazawa:2012at,Kitazawa:2011wh,Stephanov:2008qz,Stephanov:2011pb,Athanasiou:2010kw,Asakawa:2009aj,
Stephanov:2009ra}. The long-range correlation near the critical point also leads to the acceptance dependence of the cumulants~\cite{Ling:2015yau,Jiang:2015hri,Bzdak:2016sxg}. In the experiment, the non-monotonic behavior of $\kappa\sigma^2$  and the corresponding rapidity dependence of the net-proton were observed in the Au+Au collisions with the variation of the collision energy ~\cite{STAR:2013gus,Luo:2015ewa,Thader:2016gpa,STAR:2020tga,STAR:2021iop}, indicating the existence of the critical point. Recently, it has been realized that the dynamical critical fluctuations play an essential role in the evolving QGP near the critical point, explaining the conflict sign of $S\sigma$
between the prediction of the static critical fluctuations and the corresponding experimental measurements.
It was also found that the critical slowing down effect significantly influences the behavior of cumulants, which even reverses their signs compared to the equilibrium ones~\cite{Berdnikov:1999ph,Nonaka:2004pg,Mukherjee:2015swa,Jiang:2017mji}. As the system is driven out of equilibrium, the fluctuations do not have enough time to develop and the characteristic scales are "frozen", leading to the Kibble-Zurek (KZ) scaling of the cumulants~\cite{Mukherjee:2016kyu,Wu:2018twy,Wu:2019qfz,Akamatsu:2018vjr}. For a quantitative comparison with the experimental measurements, a more realistic description is required and intensive studies on the dynamical models are under development ~\cite{Nahrgang:2011mg,Herold:2016uvv,Paech:2003fe,Kapusta:2011gt,An:2019osr,An:2019csj,Murase:2016rhl,Akamatsu:2016llw,Martinez:2018wia,Stephanov:2017ghc,Du:2020bxp,Rajagopal:2021doy,Pradeep:2022mkf,Sakaida:2017rtj,Schaefer:2022bfm,Nahrgang:2018afz,Nahrgang:2020yxm,Pihan:2022xcl,Pradeep:2022eil}, please see Refs.~\cite{Bzdak:2019pkr,An:2021wof,Wu:2021xgu} for the recent review.

However, most of the dynamical model calculations within model A and model B  implement a trajectory with a fixed chemical potential and assume a Hubble-like expansion
to study the critical fluctuations near the QCD critical point~\cite{Berdnikov:1999ph,Nonaka:2004pg,Mukherjee:2015swa,Jiang:2017mji,Nahrgang:2018afz,Pihan:2022xcl,Nahrgang:2020yxm}.
In this work, we study the dynamics of the order parameter field near the QCD critical point with a realistic QGP trajectory and the cooling rate obtained from hydrodynamic simulations.
We find that the critical slowing down effects associated with the cooling rate of hydrodynamics are unexpectedly large, leading to a dramatic suppression of higher-order cumulants.
Besides implementing the realistic hydrodynamic trajectory, we also tune the location of the critical point in the potential of the Langevin equation of model A to study the interplay between two competing factors: as the evolving system approaches the critical point on the phase diagram, the increasing correlation length leads to an enhancement of fluctuations, but also strengthens the critical slowing down effect, which in turn suppresses the critical fluctuations. As the system gets very close to the critical point, the suppression from the critical slowing down effects overcomes the increase of the critical fluctuations. Therefore, the maximum of the fluctuations does not necessarily correspond to the evolving trajectory closest to the critical point.

\section{Model and Setups}\label{Model and Setups}

For the dynamical models near the critical point, it is essential to determine the dynamical universality class~\cite{Hohenberg:1977ym}. It has been argued that the evolving hot QCD system belongs to model H~\cite{Son:2004iv}, which describes the dynamics of the system with the conserved order parameter, the conserved momentum density, and the Poisson bracket between them. For the numerical simulations,  model H is still too complicated to be implemented. In this paper, we start with a simplified model, model A in Hohenberg's classification, which focuses on the dynamics of the non-conserved order parameter field $\sigma$~\cite{Hohenberg:1977ym}, together with the implementation of hydrodynamics to provide the heat bath for the order parameter field. As the first step of study, model A presents a reasonable description of the dynamics near the QCD critical point, including critical slowing down effects\cite{Mukherjee:2015swa}, dynamical critical scaling~\cite{Schaefer:2022bfm,Mukherjee:2016kyu,Wu:2018twy}.

Within model A, the evolution of the $\sigma$ field is described by the Langevin equation\cite{Wu:2018twy}:
\begin{equation}
    \frac{\partial \sigma(\textbf{x},\tau)}{\partial \tau} = -\frac{1}{m_\sigma^2 \tau_{\mbox{\tiny eff}}}\frac{\delta U[\sigma(\textbf{x})]}{\delta\sigma(\textbf{x})} + \zeta(\textbf{x},\tau),\label{Langevin}
\end{equation}
where the noise term $\zeta(\textbf{x},\tau)$ satisfies the fluctuation-dissipation theorem:
\begin{equation}
\begin{aligned}
    \langle\zeta(\textbf{x},\tau)\rangle&=0,\\
    \langle\zeta(\textbf{x},\tau)\zeta(\textbf{x}',\tau')\rangle&=\frac{2T}{m_\sigma^2 \tau_{\mbox{\tiny eff}}}\delta^3(\textbf{x}-\textbf{x}')\delta(\tau-\tau'),
    \label{noise}
\end{aligned}
\end{equation}
and $\sigma(\textbf{x})$ is the order parameter field, $m_\sigma$ is the mass of the order parameter field, $T$ is the temperature. $\tau_{\mbox{\tiny eff}}$ is the effective relaxation time with the form of $\tau_{\mbox{\tiny eff}} = \tau_{\mbox{\tiny rel}}(\xi_{\mbox{\tiny eq}}/\xi_{\mbox{\tiny min}})^z$ with $\tau_{\mbox{\tiny rel}} = 0.05$fm in this work. $\xi_{\mbox{\tiny min}}$ is the correlation length at the edge of the critical region and $\xi_{\mbox{\tiny eq}}$ is the equilibrium correlation length of the system. According to~\cite{Wu:2018twy,Mukherjee:2015swa,Hohenberg:1977ym}, the dynamical critical exponent is set to $z=3$ as in model H in this work. 

Note that the noise term within the framework of model A is typically adopted as the white noise, as shown in Eq.(\ref{noise}), in the first step of study. We assume that all the correlation effect has been encoded in the effective potential as the first term in Eq.(\ref{Langevin}) and no additional correlation in the noise term. For the realistic description of the dynamical critical fluctuations with the spatially inhomogeneous QGP fireball, the noise can be extended to the multiplicative noise, as in Ref.\cite{Chao:2020kcf}.

$U[\sigma(\textbf{x})]$ is the effective potential that can be expanded into the powers of the order parameter field $\sigma(\textbf{x})$ near the critical point:
\begin{equation}
\begin{aligned}
    U[\sigma(\textbf{x})] =\int d^3\textbf{x}
    &\frac{1}{2}[\nabla\sigma(\textbf{x})]^2
    + \frac{1}{2}m_{\sigma}^2[\sigma(\textbf{x})-\sigma_0]^2\\
    &+ \frac{\lambda_3}{3}[\sigma(\textbf{x})-\sigma_0]^3
    + \frac{\lambda_4}{4}[\sigma(\textbf{x})-\sigma_0]^4,
    \label{potential}
\end{aligned}
\end{equation}
where $\sigma_0 = \int d^3\textbf{x}\sigma(\textbf{x})/V  $ is the equilibrium mean value of $\sigma(\textbf{x})$, $\lambda_3$ and $\lambda_4$ are the coupling coefficients of the cubic and quartic terms respectively, $m_{\sigma}$ is the mass of the order parameter field related to the correlation length $\xi_{\mbox{\tiny eq}}$ by $m_\sigma=1/\xi_{\mbox{\tiny eq}}$. The equilibrium mean value $\sigma_0$, correlation length $\xi_{\mbox{\tiny eq}}$ and coupling coefficients $\lambda_3,\lambda_4$ in the effective potential $U[\sigma(\textbf{x})]$ for the QCD system are obtained through mapping from the three-dimensional Ising model~\cite{Mukherjee:2015swa,Zinn-Justin:1999opn,Schofield:1969zz}.
In more detail, the cumulants are calculated both from the distribution function $P[\sigma] \sim \exp[-U(\sigma)/T]$ in the hot QCD system and from the parametric magnetization $M_{eq}$ in the three-dimension Ising model. By comparing the cumulants from these two approaches, the coefficients of the hot QCD systems can be expressed in the form of $
    \sigma_0 = M_0 R^\frac{1}{3} \theta,\,
    \xi^2 = (M_0/H_0)[R^\frac{4}{3}(3+2\theta^2)]^{-1},\,
    \lambda_3 = (2H_0/M_0^2)R\theta(9+\theta^2)(3-\theta^2)^{-1},\,
    \lambda_4 = (2H_0/M_0^3)R^\frac{2}{3}(27+45\theta^2-31\theta^4-\theta^6)(3-\theta^2)^{-3}.
    $
$R$ and $\theta$ are two parameters associated with the Ising model variables $(r, h)$:
\begin{equation}
\begin{aligned}
    r(R,\theta)=R(1-\theta^2),\quad h(R,\theta)=R^{\frac{5}{3}}(3\theta-2\theta^3).
    \label{Rtheta}
\end{aligned}
\end{equation}
 Meanwhile, the variables $(r, h)$ in the 3D Ising model system are associated with the variables $(T, \mu)$ in the QCD system by such a mapping:
 \begin{equation}
\begin{aligned}
      \frac{T-T_c}{\Delta T} = \frac{h}{\Delta h},   \frac{\mu-\mu_{Bc}}{\Delta \mu} = -\frac{r}{\Delta r},
      \label{Tmu}
\end{aligned}
\end{equation}
where $T_c$ is the critical temperature and $\mu_{Bc}$ is the critical chemical potential, $\Delta T$ and $\Delta \mu$ are the widths of the critical regime in the QCD phase diagram, $\Delta h$ and $\Delta r$ are the corresponding widths in the Ising model. In this work, we set $\Delta T = 20$ MeV, $\Delta \mu = 100$ MeV, $\Delta h = 2$, $\Delta r = (5/3)^{3/4}$. Note that the mapping from Ising variables to the QCD ones is non-universal, depending on the choice of the parameters, such as $H_0,M_0$ etc. Since we have little knowledge of this mapping, we treat them as free parameters and take their values following Ref.\cite{Wu:2018twy}, {\it i.e.,} $M_0=200$MeV and $H_0=M_0/5$ within reasonable parameter values. The exploration of the parameter space has been studied in Ref.\cite{Parotto:2018pwx}, in which we believe the values of these non-universal parameters do not modify the qualitative behavior of critical fluctuations.

\begin{figure}[!htpb]
\centering 
\includegraphics[width=0.4\textwidth]{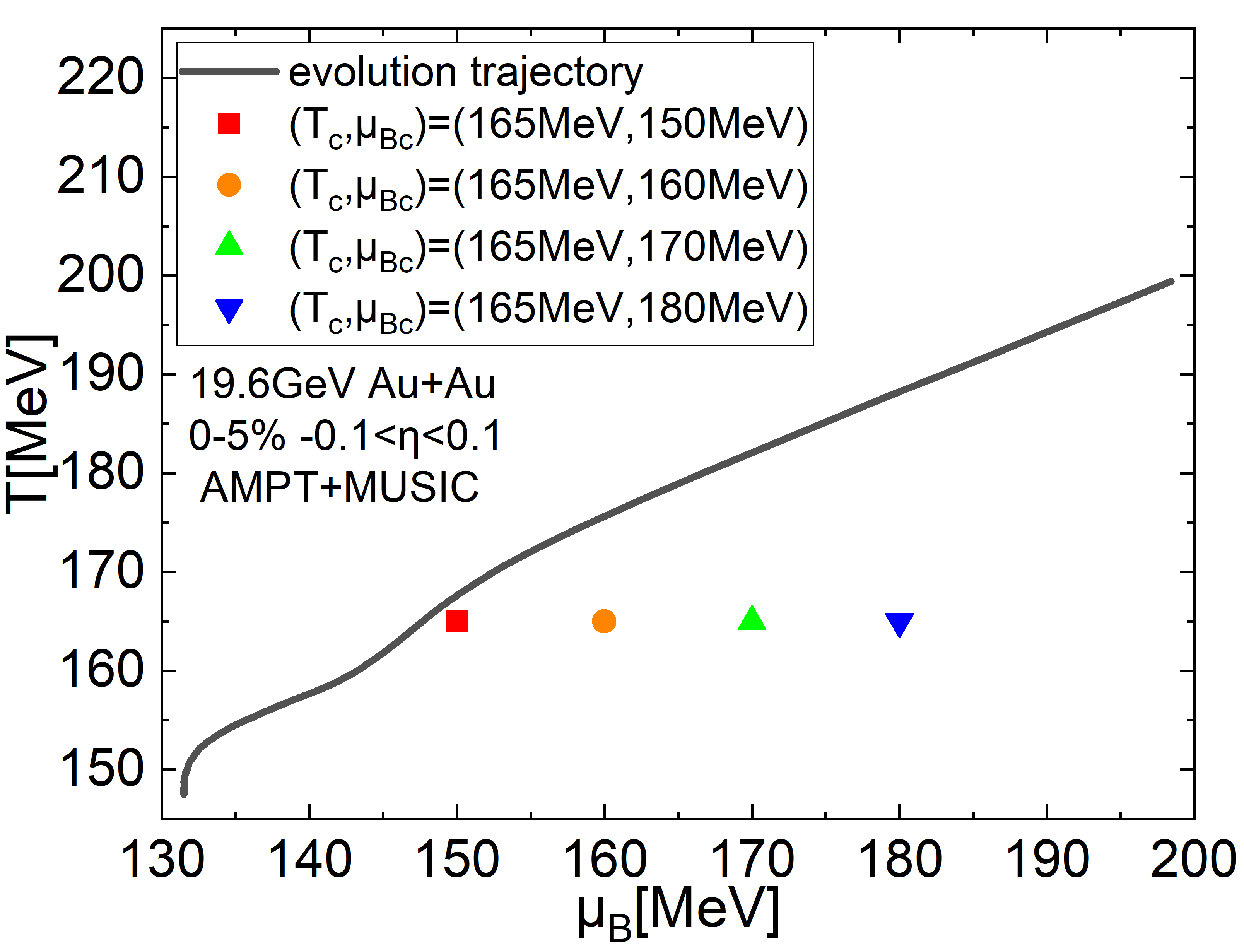} 
\caption{(Color online) The average QGP evolution trajectory on the $T,\mu$ plane, obtained from MUSIC simulations for 19.6 GeV Au+Au collisions at 0-5\% centrality. Points with different colors correspond to different locations of the critical point.}
\label{trajectory} 
\end{figure} 

\begin{figure}[!htpb]
\centering 
\includegraphics[width=0.4\textwidth]{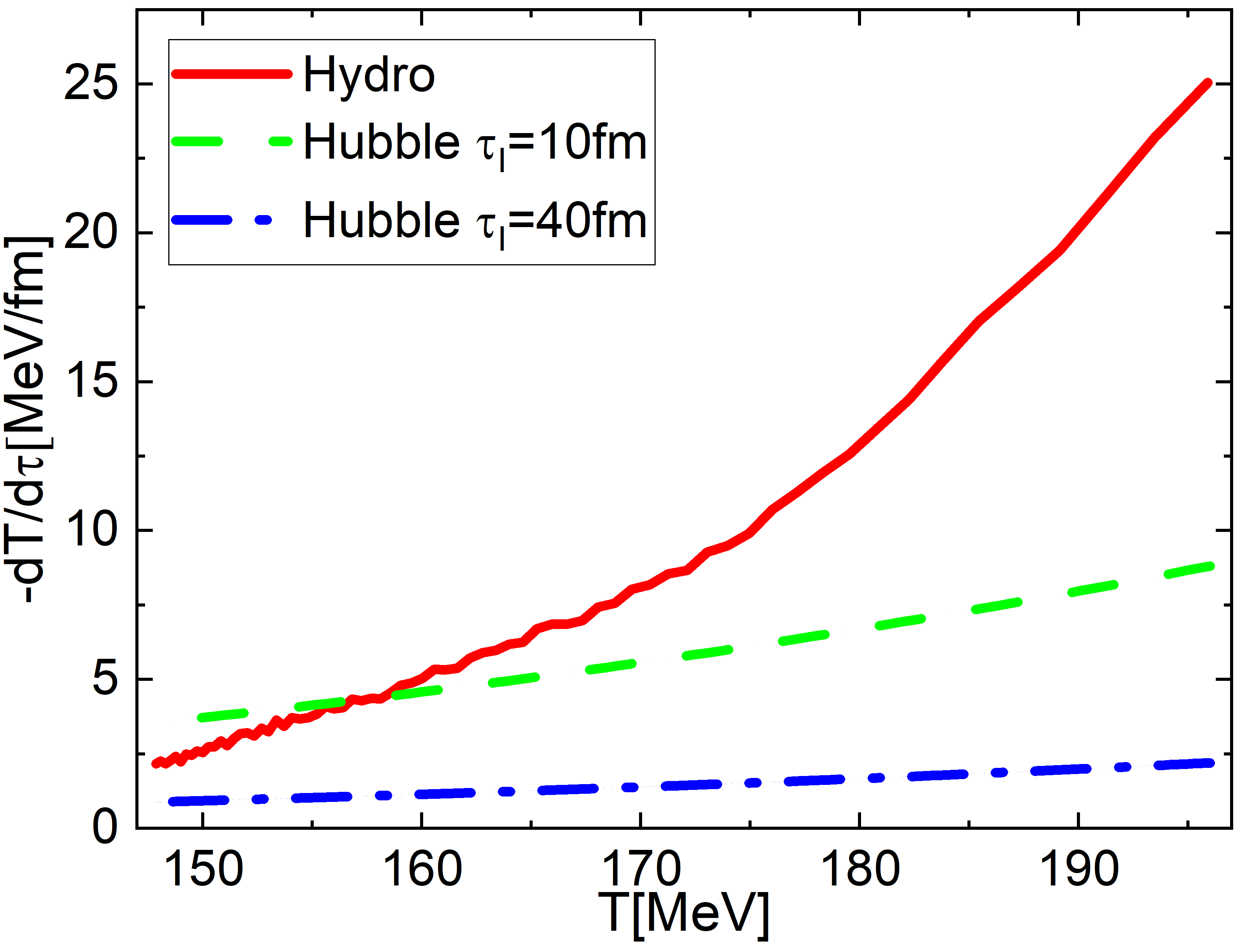} 
\caption{(Color online) A comparison of the cooling rates between hydrodynamic system and Hubble-like expansion systems.}
\label{dTdtau} 
\end{figure} 

\begin{figure*}[!htpb]
\centering 
\includegraphics[width=0.8\textwidth]{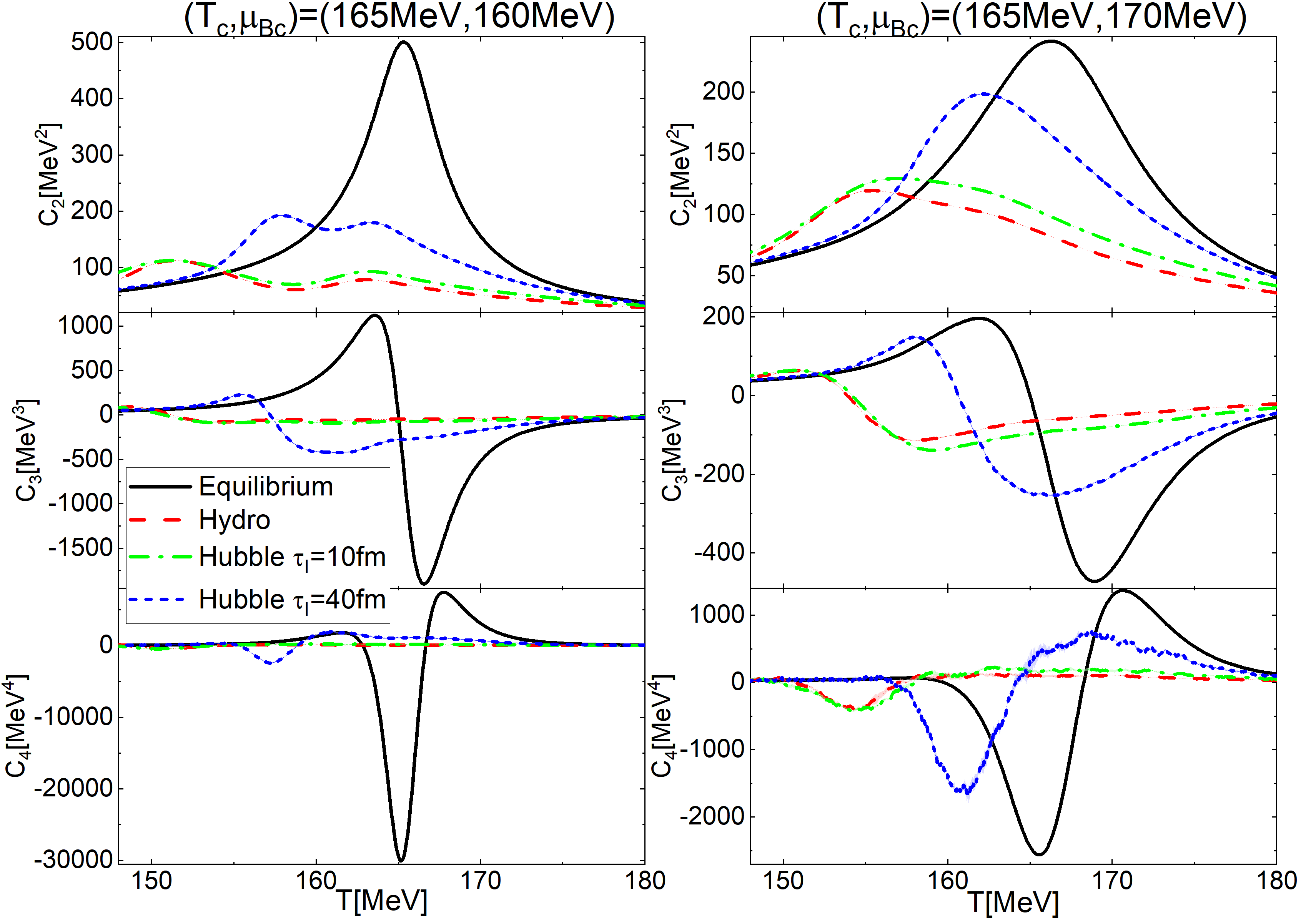} 
\caption{(Color online) The evolution of cumulants of order parameter as functions of temperature. Curves with different colors correspond to different cooling rates. Critical point is placed with $(T_c, \mu_{Bc}$ = 165 MeV, 160 MeV) and $(T_c, \mu_{Bc}$ = 165 MeV, 170 MeV) for left and right columns, respectively.}

\label{sigma1} 
\end{figure*} 

\begin{figure*}[!htpb]
\centering 
\includegraphics[width=0.8\textwidth]{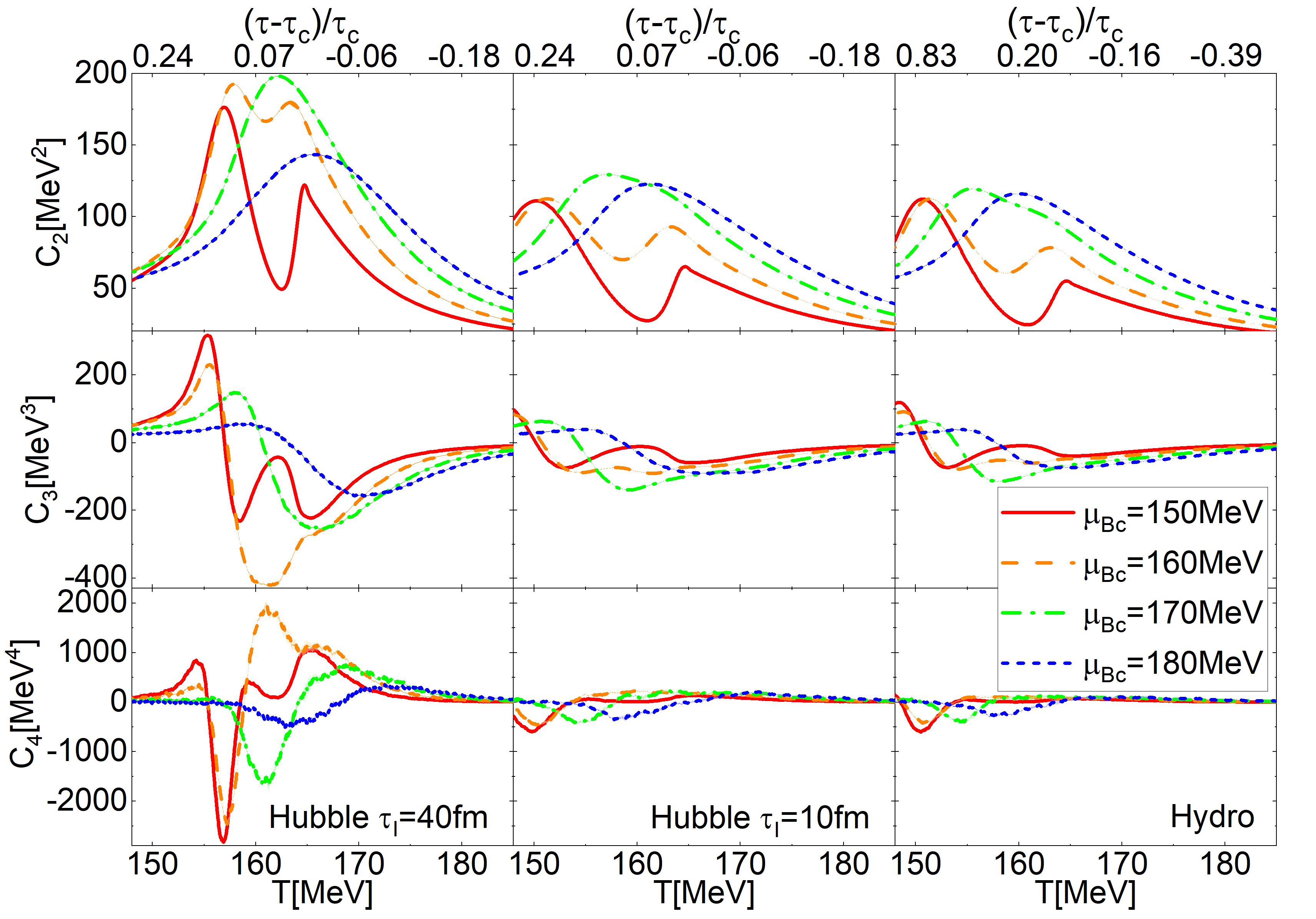} 
\caption{(Color online) The evolution of cumulants of order parameter as functions of temperature. Curves with different colors correspond to different locations of the critical point. Cooling rates originate from $\tau_I=40 \mathrm{fm}$ Hubble expansion, $\tau_I=10 \mathrm{fm}$ Hubble expansion and Hydrodynamics from left to right columns, respectively.}
\label{sigma2} 
\end{figure*} 

Thermodynamic properties of the heat bath, such as the temperature $T(\mathbf{x})$ and the baryon chemical potential $\mu(\mathbf{x})$, are treated as inputs of Eq.\eqref{Langevin}, which are obtained from hydrodynamic simulations with the assumption of local equilibrium. To simplify the numerical simulations of Eq.\eqref{Langevin}, we make an additional average for the temperature and baryon chemical potential profiles over the whole QGP fireball with the energy density as the weight: $T=\langle T(\mathbf{x})\rangle$, $\mu=\langle\mu(\mathbf{x})\rangle$. As a result, the time evolution of the fireballs created in heavy ion collisions at different collision energies and centralities is simplified as evolving trajectories on the QCD phase diagram.

In this work, the evolution profiles for the QGP fireball are generated from the 3+1-d hydrodynamics MUSIC~\cite{Shen:2014vra} with the initial profiles constructed from the transport model AMPT\cite{Lin:2004en,Xu:2016hmp,Zhao:2017yhj}. Here, we input an equation of state using the lattice simulation results, together with incorporating a critical point~\cite{Parotto:2018pwx}. To roughly fit the multiplicity, spectra and flow in 19.6 GeV Au+Au collisions, we set the parameters in numerical simulations as follows, $\tau_{0I}=0.4$fm is the starting proper time for the AMPT initial condition while $\tau_{0h}=1.2$fm is the starting time for hydrodynamics evolution, the specific shear and bulk viscosity are set as $\eta/s=0.08$, $\zeta/s=0$. We set the switching temperature as $T_{sw}=147$ MeV that transit hydrodynamic simulation to UrQMD, which is lower than the critical temperature $T_c=165$MeV for the sake of studying the critical slowing down effect. Fig.\ref{trajectory} shows the average QGP evolution trajectory on the $T,\mu$ plane, obtained from MUSIC simulations for 19.6 GeV Au+Au collisions at 0-5\% centrality.
To study the critical slowing down effects on the magnitude of the critical fluctuations, we change the location of the critical point, but keep the trajectory of the QGP fireball fixed as shown in Fig.\ref{trajectory}. In this work, we choose four locations of the critical point, which fix the critical temperature at $T_c=165$ MeV, but change the critical chemical potential as $\mu_{Bc}=150,160,170$, and $180$ MeV.

In addition to the time evolution of the temperature and chemical potential obtained from hydrodynamic simulations, we also compare with another expansion case, the Hubble-like expansion, which is described as~\cite{Mukherjee:2015swa}:
\begin{equation}
\begin{aligned}
     \frac{T}{T_I} = \left(\frac{\tau}{\tau_I}\right)^{-n_V c_s^2},
\label{Hubble}
\end{aligned}
\end{equation}
where the initial temperature is set to be $T_I = 196$ MeV, and the speed of sound is set to be $c_s^2 = 0.15$. $n_V$ denotes the dimension of the expansion system, and we choose $n_V = 3$ for the three-dimensional Hubble-like expansion. The initial times are set as $\tau_I = 10, 40$ fm for two different cooling rates of Hubble-like expansion. To obtain the Hubble-like expansion~Eq.\eqref{Hubble}, the volume is assumed as $V/V_I = (\tau/\tau_I)^{n_V}$ and the total entropy is approximately conserved during the evolution. Hence the entropy density evolves as $s/s_I = (\tau/\tau_I)^{-n_V}$, together with the thermodynamic relationship $s \propto T^{1/c_s^2}$, then Eq.(\ref{Hubble}) is obtained. Here, we also set the system to evolve along the trajectory obtained from hydrodynamic as shown in Fig.\ref{trajectory}. As the temperature decreases as in Eq.\eqref{Hubble}, the associated chemical potential is set to the corresponding value obtained from the trajectory. Fig.\ref{dTdtau} compares the cooling rates of these three systems: $(T(\tau),\mu(\tau))$ from hydrodynamics, Hubble-like expansion with $\tau_I=10$ and 40 fm. The cooling rate of the hydrodynamic system is faster than that of the two Hubble-like expansion systems, and we tune $\tau_I=10$fm for the similar expanding system near the QCD critical point and $\tau_I=40$fm for a slower one.

With the above settings, the Langevin equation Eq.(\ref{Langevin}) is solved numerically with the discretization in a cubic box with lattice spacing $\Delta x=1$fm and volume $V=10^3$fm$^3$. For the increment in each temporal step, we choose $\Delta t=0.01$fm. The initial profile of the $\sigma$ field is constructed from the distribution function: $P[\sigma] \sim \exp[-U(\sigma)/T]$. The discretization of the noise term induces the lattice-spacing dependence with a cut-off. In principle, the cut-off can be absorbed into the redefined transport coefficient and effective potential by the renormalization scheme~\cite{Cassol-Seewald:2007oak}. However, it is hard to implement in the expanding system, which has not been achieved in numerical simulations. For simplicity, we treat the noise in Eq.(\ref{noise}) as uniform in spatial dimension but random in temporal dimension to avoid the lattice-spacing dependence, following Refs.\cite{Herold:2016uvv,Jiang:2021fun,Jiang:2021gsw,Wu:2018twy}. Eventually, we evolve the Langevin equation independently for each event with the event number up to 4.5 million.

The cumulants of the $\sigma$ field can be calculated as:
\begin{equation}
\begin{aligned}\label{Eq:cumulants}
    C_1 &= \langle\sigma\rangle,\\
    C_2 &= \langle\sigma^2\rangle-\langle\sigma\rangle^2,\\
    C_3 &= \langle\sigma^3\rangle-3\langle\sigma^2\rangle\langle\sigma\rangle+2\langle\sigma\rangle^3,\\
    C_4 &= \langle\sigma^4\rangle-4\langle\sigma^3\rangle\langle\sigma\rangle-3\langle\sigma^2\rangle^2+12\langle\sigma^2\rangle\langle\sigma\rangle^2-6\langle\sigma\rangle^4,
\end{aligned}
\end{equation}
where $\sigma$ denotes the spatial average of the order parameter field and $\langle ... \rangle$ represents the event average. In Sec.\ref{Result and Discussion}, we also need to calculate the equilibrium cumulants of the $\sigma$ filed for comparison, which are also obtained from Eq.\eqref{Eq:cumulants}, but with $\langle ... \rangle$ denoting the average with the distribution function $P[\sigma]\sim\exp[-U(\sigma)/T]$. The corresponding effective potential $U[\sigma]$ is calculated with the equilibrium values of $\sigma_0,m_\sigma,\lambda_3$ and $\lambda_4$ at each point of the trajectory.

\section{Result and Discussion}\label{Result and Discussion}

Critical slowing down effects have been found near the QCD critical point within model A~\cite{Mukherjee:2015swa,Berdnikov:1999ph,Nonaka:2004pg} and B~\cite{Nahrgang:2018afz,Pihan:2022xcl,Nahrgang:2020yxm}, where the evolution of temperature is parametrized with the cooling rate of Hubble-like expansion. In this work, we focus on analyzing the critical slowing down effects with the hydrodynamic expansion, using the temperature and chemical potential profiles along the hydrodynamic trajectories as shown in Fig.\ref{trajectory}. To obtain qualitative picture, Fig.\ref{sigma1} also compares the time evolution of the cumulants, using the hydrodynamic cooling rate and the Hubble-like cooling rates with different $\tau_I$.  Since we have no knowledge of the location of the critical point, we treat it as a free parameter in the evolving equation Eq.(\ref{Langevin}). Here we choose two locations of the critical point ($T_c=165\ \mathrm{MeV}, \  \mu_{Bc}=160\ \mathrm{MeV}$) and ($T_c=165\ \mathrm{MeV}, \  \mu_{Bc}=170\ \mathrm{MeV}$). As shown in Fig.\ref{sigma1}, the non-equilibrium cumulants(color curves) show memory effects, which have a similar trend as the equilibrium ones(black curves), but reach the maximum(or minimum) at a later time\footnote{One can also see the dip structure in $C_2$ in the left panel of Fig.3, which is caused by the non-Gaussian term in Eq.(\ref{potential}). Such non-Gaussian effect is mainly determined by $\xi^3/V$, which leads to a more obvious dip structure of $C_2$ for a system closer to the critical point.}. Due to the critical slowing down effects, the magnitudes of various non-equilibrium cumulants are suppressed compared with the equilibrium ones, and the suppression increases with a larger cooling rate.
For the non-equilibrium cumulants with the cooling rate obtained from the realistic hydrodynamic simulation, dramatic suppressions are observed,  and the fourth-order cumulant $C_4$ even becomes flat as a function of $T$ compared to the equilibrium one. Compared to the right column, the left column with the critical point set to $(T_c,\mu_{Bc})=(165 \mbox{MeV},160 \mbox{MeV})$ shows an enhancement of the equilibrium cumulants as the system get closer to the critical point with an increasing correlation length of the $\sigma$ field. However, the corresponding non-equilibrium cumulants are also largely suppressed due to the larger critical slowing down effects.

To further study the critical slowing down effects, we change the location of the critical point in the potential of the Langevin equation (Eq.\eqref{Langevin}), but keep the hydrodynamic evolution trajectory fixed, as shown in Fig.\ref{trajectory}. Fig.\ref{sigma2} plots the cumulants of the $\sigma$ field, simulated by the Langevin equation, using the four different locations of the critical point. From left to right columns, we use the same evolution trajectory for the temperature and chemical potential profiles but with different cooling rates described by the Hubble-like expansion Eq.\eqref{Hubble} and obtained from the realistic hydrodynamic simulation. Even for a slow Hubble expansion system
with $\tau_I=40 \mathrm{fm}$ (left column), the magnitudes of the different orders of the cumulants do not monotonically increase as the system approaches the critical point ($\mu_{Bc}=180$ MeV to 150MeV).  For the faster expanding systems (middle and right column in Fig.\ref{sigma2}), this non-monotonicity in terms of the distance to the critical point becomes more obvious because of larger critical slowing down effects.

\begin{figure}[!htpb]

\centering 

\includegraphics[width=0.4\textwidth]{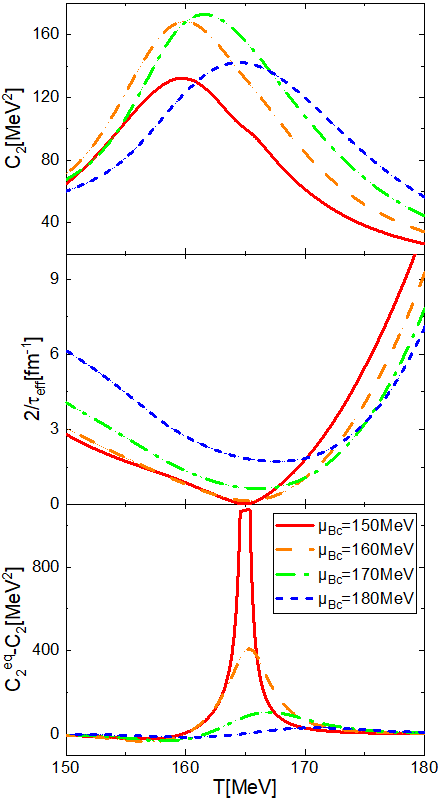} 

\caption{(Color online) The evolution of the second order cumulants(top), the inverse of the effective relaxation time $2/\tau_{\mbox{\tiny eff}}$ (middle) and the difference between the equilibrium and non-equilibrium cumulants $C^{\mbox{\tiny eq}}_2-C_2$ (bottom), calculated from a simplified equation Eq.\eqref{sim} with different locations of the critical point. Curves with different colors correspond to different locations of the critical point.}
\label{evo} 

\end{figure} 


To understand the essential mechanism of this non-monotonicity in terms of the distance to critical point, we check the evolution equation of the second-order cumulants:
\begin{equation}
\begin{aligned}
     \frac{\partial C_2}{\partial \tau} = -\frac{2}{\tau_{\mbox{\tiny eff}}}\left[C_2 - C_2^{\mbox{\tiny eq}}\right].
\label{sim}
\end{aligned}
\end{equation}
It is a simplified equation of Eq.\eqref{Langevin} that neglects the third- and fourth-order terms~\cite{Wu:2018twy}
, with $\tau_{\mbox{\tiny eff}} = \tau_{\mbox{\tiny rel}}(\xi_{\mbox{\tiny eq}}/\xi_{\mbox{\tiny min}})^z$, $\tau_{\mbox{\tiny rel}} = 0.05$fm.
The top panel of Fig.\ref{evo} plots the evolution of the second-order cumulant $C_2$, calculated from the simplified equation Eq.\eqref{sim} with a realistic trajectory and cooling rate obtained from hydrodynamics simulation. Just like the case from full Langevin equation simulation shown in the upper panels of Fig.\ref{sigma2}, the peak value of $C_2$ here first increases and then decreases as the system approaches the critical point from $\mu_{Bc}=180$ MeV to 150 MeV. To illustrate how the critical slowing down effects play a role in the evolution of $C_2$, we also plot the evolution of $2/\tau_{\mbox{\tiny eff}}$ and $C^{\mbox{\tiny eq}}_2-C_2$ in the middle and bottom panels. Here, $C^{\mbox{\tiny eq}}_2-C_2$ represents the difference between the non-equilibrium cumulant $C_2$ and the equilibrium one $C^{\mbox{\tiny eq}}_2$. As expected,  $C^{\mbox{\tiny eq}}_2-C_2$ increases as the system approaches the critical point from $\mu_{Bc}=180$ MeV to 150 MeV.
However, the critical slowing down effects also increase dramatically. As $2/\tau_{\mbox{\tiny eff}}$ approaches zero value, it takes an infinitely long time for $C_2$ to catch up with the values of $C^{\mbox{\tiny eq}}_2$. As a result, the enhanced critical slowing down effects strongly suppress the critical fluctuations, even when the system is very close to the critical point with a dramatically increased correlation length. This also leads to the non-monotonic behavior of $C_2$ as the system approaches the critical point. Note that we analyze the critical slowing down effects of second-order cumulant in Eq.\eqref{sim} with the case of the fastest cooling rate in this work, the one with the hydrodynamic system. The case with Hubble-like expansion is much slower but the argument is applicable as well.

\section{Conclusion and Outlook}\label{conclusion}

Within the framework of model A that evolves the non-conserved order parameter field $\sigma$, we study
the interplay between the critical fluctuations and the critical slowing down effects, using the QGP evolution trajectory and cooling rate obtained from hydrodynamics. To study the critical slowing down effects, we also change the location of the critical point in the potential of the Langevin equation and compare the simulations with different cooling rates from hydrodynamics and the Hubble-like expansion. As discovered by early studies, we also found that the critical slowing downing effects suppress the critical fluctuations, which even reverse the sign of higher order cumulants for the Hubble-like expansion system with a small cooling rate. However, our comparison simulations show that the cooling rate from a realistic hydrodynamic system is pretty large which leads to a huge suppression of the higher-order cumulants of the order parameter field ({\it e.g.,} $C_4$). Furthermore, as the system gets very close to the critical point with largely enhanced correlation length, the dramatically increased critical slowing down effects lead to a large suppression of the critical fluctuations ({\it i.e.,} $C_2$), even for the system that is very close to the critical point.

Finally, we would like to point out that this study of critical fluctuation implements a simplified model A that evolves only the non-conserved order parameter field. Here, the evolution of the hydrodynamics is decoupled from the evolution of the order parameter field. Its temperature and chemical potential profiles along the averaged evolution trajectory are treated only as the inputs to model A. To get a more insightful interpretation of the RHIC-BES observables, more realistic models are required for future studies. For example, for the inhomogeneous fireball created in relativistic heavy-ion collisions, the dynamics of order parameter that couples with hydrodynamics in a non-trivial way requires more theoretical study. Although the simulation of the dynamics for the conserved baryon density near the QCD critical point ({\it i.e.}, model B) has been carried out in Ref.~\cite{Nahrgang:2018afz,Pihan:2022xcl,Nahrgang:2020yxm}, it should be extended with the inhomogeneous fireball background. In addition, for the comparison with experimental data, it is also necessary to study  the coupling between the order parameter field  and final protons in the freeze-out process within model A~\cite{Pradeep:2022mkf,Jiang:2015hri}.

\section*{Acknowledgements}

This work is supported by the NSFC under grant No. 12247107, 12075007 and No.11947236 as well as the China Postdoctoral Science Foundation under Grant No. 2020M680184. We also acknowledge the extensive computing resources provided by the Supercomputing Center of Chinese Academy of Science (SCCAS), Beijing Super Cloud Computing Center (BSCC), Tianhe-1A from the National Supercomputing Center in
Tianjin, China and the High-performance Computing Platform of Peking University.

\end{document}